\documentclass[12pt, a4paper]{article}
\usepackage{graphicx}
\begin{document}
\setcounter {equation}0
\noindent
{\large \bf How precisely can the difference method determine the
\mbox{\boldmath{$\pi N\! N$}} coupling constant?}\\
B.~Loiseau$^{a,}${\footnote{loiseau@in2p3.fr} and
\vspace{3mm}
T.E.O.~Ericson$^{b,c,}${\footnote{torleif.ericson@cern.ch}} \\
{\small 
$^a$LPNHE/LPTPE, Universit\'{e} P. \& M. Curie,
4 Place Jussieu, 75252 Paris Cedex 05, France\\
$^b$CERN, CH-1211 Geneva 23, Switzerland\\
$^c$TSL, Box 533, S-75121 Uppsala, Sweden}


\begin{abstract}

The Coulomb-like backward peak of the  np scattering differential cross
section is due to one-pion exchange. Extrapolation to the pion pole  
of precise data should allow to obtain the value of the  charged 
$\pi N\! N$ coupling constant. This was classically attempted by the use of a
smooth physical function, the Chew function, built from the cross section.
To improve accuracy of such an extrapolation one
has introduced a difference method. It consists of  extrapolating
the difference between the Chew function based on experimental data and 
that built from a model where the $\pi N\! N$ coupling is exactly known.
Here we cross-check 
to which precision can work this novel extrapolation method by
applying it to differences between models and between data and models.
With good reference models and for the 162 MeV np Uppsala 
single energy precise data with a
normalisation error of 2.3~\%,
the value of the charged $\pi N\! N$ coupling constant is 
obtained with an accuracy close to 1.8~\%.

\end{abstract}

\section{Introduction}
\label{intro}

It is of crucial importance to know the precise value of the $\pi N\! N$
coupling constant, both in Nuclear and in Particle physics. Together with the
pion mass this coupling  scales the nuclear interaction. Through the
Goldberger-Treiman relation~\cite{Goldberger58} it tests chiral symmetry.
From this last relation one expects an accuracy in its value of about
1\% as has been discussed in details in Ref.~\cite{Rahm98}. Some further
recent considerations on this relation and its implications for the 
$\pi N\! N$ coupling can be found, for instance, in Ref.~\cite{Goity99}.

In the 1980,  the $\pi N\! N$ coupling constant was believed to be
well known. The analysis of $\pi^{\pm} p$ scattering data~\cite{Koch80} gives
a value of $14.28 \pm 0.18$ for the  charged pion coupling constant.
Forward dispersion relation analysis of pp scattering 
data~\cite{Kroll81} led to  $g^{2}_{\pi^{0}}/4\pi = 14.52 \pm 0.40$ 
for the  neutral pion coupling constant.
The Nijmegen group~\cite{Bergervoet90,Klomp91,Stoks93a}, in the 1990's and
on the basis of  energy-dependent partial-wave analyses
(PWA) of nucleon-nucleon  ($N\! N$) scattering data, found
smaller values. They obtained
$g^{2}_{\pi^{0}}/4\pi  = 13.47 \pm 0.11$ and
$g^{2}_{\pi^{\pm}}/4\pi  = 13.58 \pm 0.05$. These values were confirmed in
their
more recent $NN$ PWA analyses~\cite {SWA97}.
The Virginia Polytechnic Institute (VPI) group~\cite{Arn90,Arn94a} from
analysis of both  $\pi^{\pm} N$ and $N\! N$ data has obtained
also low values around $g^2_{\pi}/4\pi = 13.7$.
From a PWA for the $\pi^+$p reaction   a value of 13.45(14) was recently
obtained~\cite{TIM97}. The very recent $\pi$N~\cite{PAV99} and
pion-photo-production~\cite{Arn99} PWA VPI analyses give values of
$13.73 \pm 0.10$  and $14.00 \pm 0.13$ respectively. Let us mention that
some charge dependence has been also considered~\cite{Meissner97,Machleidt99}.
All the determinations which rely  on  the 
analysis of large data bases from a great number of experiments,
with some of the data rejected according to certain criteria, have a very good
statistical accuracy. It is however difficult to assert them a
clear systematic uncertainty.

\begin{table}[ht]
\begin{center}
\caption {Some $\pi $NN coupling constants with their source.}
\protect\label{tab:coupling constants}
\vspace{3mm}

\begin{tabular}{|l|l|l|l|}
\hline  
  Reference                        &
   Year             &Source &   $g^2_{\pi NN}/4\pi$\\

\hline
 Karlsruhe-Helsinki \cite{Koch80} &   1980 &$\pi$p (PWA, dispersion relations)
 &14.28(18)\\
Kroll et al. \cite {Kroll81}&1981&pp (forward dispersion relations)
&14.52(40)\\
\hline
Nijmegen \cite {Stoks93a}&  1993& pp, np (PWA) &13.58(5)\\
VPI \cite{Arn94a}&  1994& pp, np (PWA)&   13.7\\
Nijmegen \cite {SWA97}&1997 &pp, np (PWA) & 13.54(5)\\
Timmermans \cite {TIM97}&1997&$\pi^+$ p (PWA) &13.45(14)\\
VPI \cite {PAV99} & 1999 & $\pi$ p (PWA, dispersion relations) & 13.73(7)\\
VPI \cite{Arn99}&  1999& $\gamma$ p $\to \pi$N (PWA)&   14.00(13)\\
\hline
VPI \cite {Arn94b}  &1994   &GMO,  $\pi$p&13.75 (15)\\
 Ericson et al. \cite{ERIC99}& 1999 &GMO, $\pi ^{\pm }$p &14.17(17)\\
\hline
Uppsala \cite{Rahm98}&1998&np$\to $pn (difference method)
&14.52(26) \\
PSI \cite{FRA99}&1999&np$\to $pn (Chew/Conformal mapping)
&13.84(43) \\
Present work &1999&np$\to $pn (difference method)
 &14.46(35)\\
\hline 
\end{tabular}
\end{center}
\end{table}

A more direct determination is the use of the Goldberger-Miyazawa-Oehme (GMO)
sum-rule~\cite{GOL55} which, in principle, depends directly
on physical observables. This was applied in particular 
in Ref.~\cite{Arn94b} giving a value of 
$g^{2}_{\pi^{\pm}}/4\pi  = 13.75\pm 0.15$ and very recently in
Ref.~\cite{ERIC99} leading to $g^{2}_{\pi^{\pm}}/4\pi  = 14.17\pm 0.17$.
Another direct method is based on the
extrapolation to the pion pole of precise data on single-energy backward
differential np cross sections. Both determinations allow a systematic
discussion of statistical and systematic uncertainties.
The use of the recent backward np Uppsala data at
162 MeV~\cite{Rahm98,Olsson99}
and of a novel
extrapolation method, the difference method, gives
$g^{2}_{\pi^{\pm}}/4\pi  = 14.52\pm 0.26$
\cite{Rahm98,Eri95}. Fairly new analysis of the recent PSI backward np data,
with the classical Chew extrapolation and a conformal mapping method leads to
$g^{2}_{\pi^{\pm}}/4\pi  = 13.84\pm 0.43$~\cite{FRA99}.
A summary of values for the coupling constant
is
given in Table~1. 
 It can be seen that between
the smallest and largest value there is a discrepancy of about 7\%.
The dispersion of the different $g^{2}_{\pi^{\pm}}/4\pi $ values can also 
be judged from Fig.~\ref{fig:gpin}.

\begin{figure}[htbp]
\begin{center}
\includegraphics[angle=90,width=12cm]{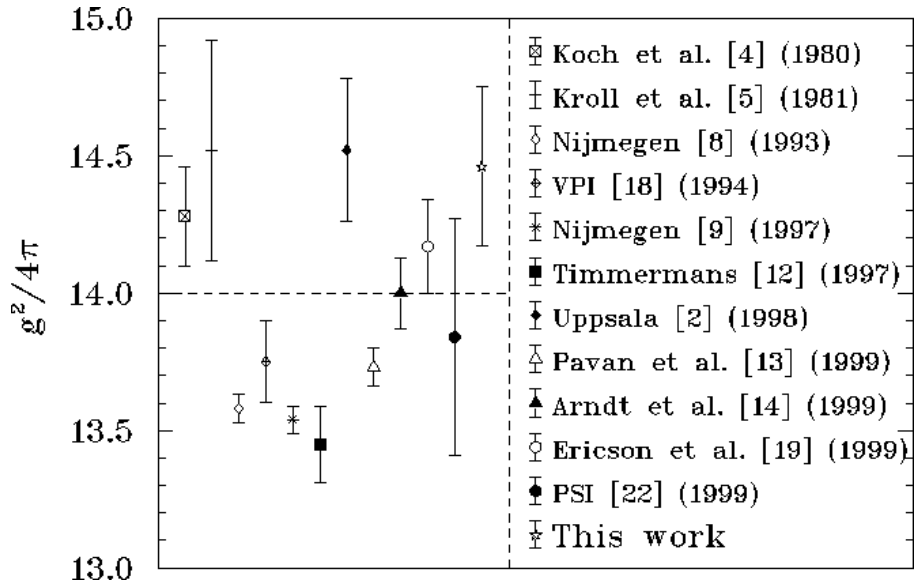}
\caption{Some $\pi NN$ coupling constants as found over the last 20 years}
\label{fig:gpin}
\end{center}
\end{figure}

Let us recall that, when using the np backward data,
the 
experimental normalisation of the cross section is very important to 
the sensitivity. 
It is both the {\em shape} of the angular distribution at the most 
backward angles, and the {\em absolute normalisation} of the data, that 
are of crucial importance~\cite{Rahm98,Olsson99,Eri95,Blomgren99}.
The Nijmegen group has strongly
criticised the Uppsala data at 162 MeV and its extrapolated result via the
difference method~\cite{SWA97,REN98a}. In particular in Ref.~\cite{SWA97} it
is claimed that there is a very strong model dependence of the difference
method. It is the purpose of the present study to check the accuracy of the
difference method and in particular, when applied to the precise 
162 MeV Uppsala data we shall demonstrate, by choosing
different models, that the model dependence is small and that for
good reference models this method leads to a precision
smaller than 2\%.

The evidence that the backward peak of the np angular distribution is
dominated by the one-pion exchange will be discussed in
Sect.~\ref{pion_exchange}.
 The determination of the 
$\pi N\! N$ coupling constant through the extrapolation to the pion pole 
from models and np data is studied in 
Sect.~\ref{extrap_pionpole}, and some conclusions are given 
in Sect.~\ref{concl}.

\section{Evidence for the one-pion exchange}
\label{pion_exchange}

\begin{figure}[htbp]
\begin{center}
\includegraphics
[angle=90,width=11cm]{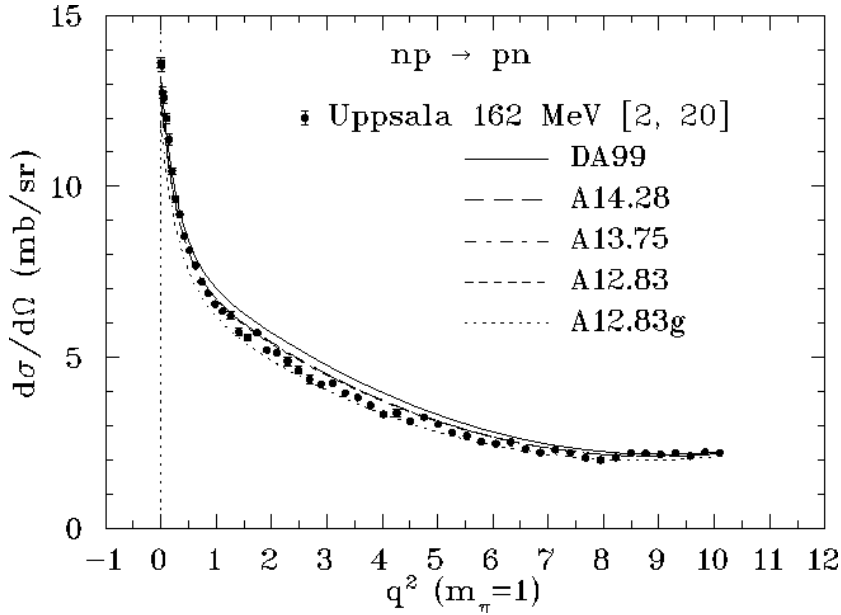}
\caption{Coulomb-like behaviour of the Svedberg Laboratory np CEX data at
162 MeV~\protect\cite{Rahm98,Olsson99} and prediction of the 'DA99'
model and of other reference models considered here~\protect\cite{Arndt99b}.}
\label{fig:162da99}
\end{center}
\end{figure}

It was early realized that the one-pion exchange (OPE) contributes
importantly to the np charge exchange (CEX) at small momentum transfer. In
Fig.~\ref{fig:162da99}
we have plotted the recent np CEX experimental
differential cross section data measured at the Svedberg Laboratory at 162
MeV~\cite{Rahm98,Olsson99} as a function of $q^2$, square of the
momentum transfer
of the neutron to the proton. It can be seen that there is a strong
peak at very small $q^2$. Note that in Fig.~\ref{fig:162da99} $q^2$ is
expressed in units of m$_\pi$, the charged pion mass.
This differential cross section has a "Coulomb like
behaviour", it goes to $\infty$, not at $q^2=0$ as in the photon-exchange case,
but at $q^2=-m^2_{\pi}$, which corresponds to the pion pole of the OPE. The
presence of this OPE pole very close to the physical region led Chew in 1958
to suggest a model-independent extrapolation to this pole which would allow
to determine the $\pi N\! N$ coupling constant~\cite{Chew58,Cziffra59}. We 
here refer, for the interested reader, to the very detailed and well
documented discussion concerning the OPE given in Ref.~\cite{Rahm98}. We
shall now recall the different methods of extrapolation which were considered
in that same reference.

\section{Extrapolation to the pion pole}
\label{extrap_pionpole}

\begin{figure} [htbp]
\begin{center}
\includegraphics
[angle=90,width=10cm]{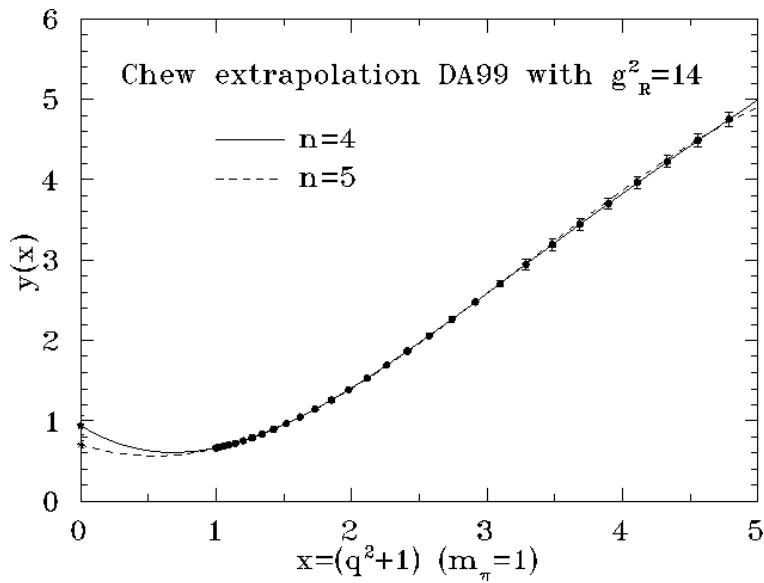}
\caption{Extrapolations to the pion pole 
at 162 MeV of the Chew function $y(q^2)$ of the model
DA99~\protect\cite{Arndt99b} for $n$-term
polynomial fits and for the reduced 
range $0 < q^2 < 4\ m_{\pi}^2$ of the 162 MeV
np CEX Uppsala data. Here the errors of these experimental data have been
assigned to the model DA99.}
\label{fig:Chew}
\end{center}
\end{figure}

\subsection{Methods of extrapolation}
\label{extrap_meth}
The  idea to extrapolate to the pion pole is to study a 
smooth physical function, the Chew function, built by multiplying the cross 
section by $(q^2+m_{\pi}^2)^2$. This removes the pole term and 
the extrapolation can be made  more safely and controllably. This function
can be then fitted to the data in the physical region by a
polynomial and extrapolated to the pole. One considers,
  \begin{equation} y(x) =
\frac{(4\pi)^2 sx^2}{m_{\pi}^4g_R^4} 
       \frac{{\rm d}\sigma}{{\rm d}\Omega}(x) = 
       \sum _{{\rm i=0}}^{{\rm n-1}}a_ix^i.
\label{eq:Chew}
\end{equation}
Here $x = q^2+m_{\pi}^2$ and $s$ is the square of the total energy. At 
the pion pole $x=0$ and
\begin{equation}
y(0) \equiv a_0 \equiv g^4_{\pi^\pm}/g_R^4
\label{eq:poleChew}
\end{equation}
where the pseudoscalar coupling constant $g^2_{\pi^\pm}/4\pi \simeq 14$. 
The quantity $g_R^2$ is a reference scale for the coupling chosen for 
convenience. The model-independent 
extrapolation requires accurate data with absolute normalisation of the 
differential cross section. If the experimental differential cross section is 
incorrectly normalised by a factor $N$, the extrapolation determines 
$\sqrt{N}g^2_{\pi^\pm}/4\pi$. This is one of the most important sources of 
uncertainty when extrapolating the data. The Chew method which has been the
most used in the past requires at least 5 terms in the $q^2$
expansion~\cite{Rahm98}.

\begin{figure}[htb]
\begin{center}
\includegraphics
[angle=90,width=11cm]{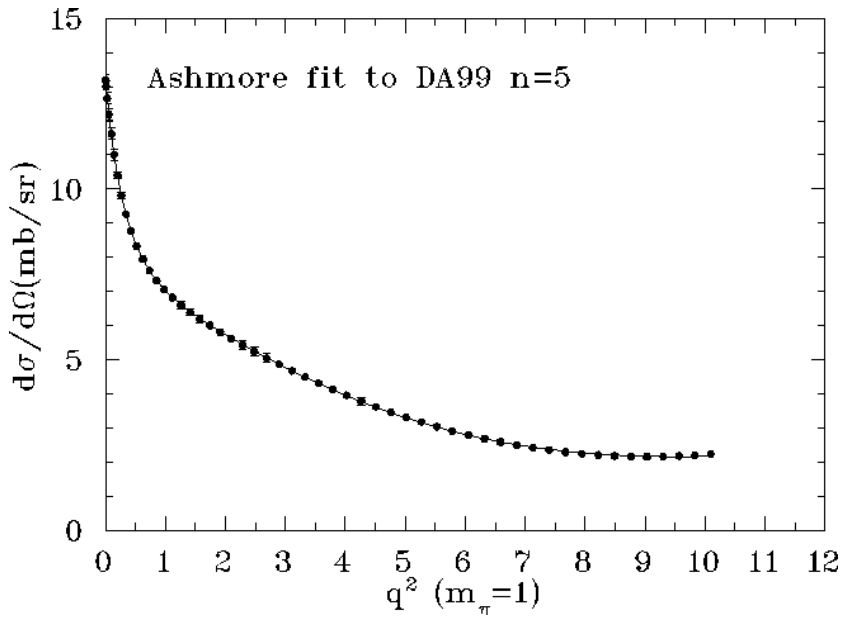}
\caption{5 terms Ashmore fit of the 
 162 MeV  pseudo-data `DA99'~\protect\cite{Arndt99b} 
 in the full range $0 < q^2 < 10.1\ m_{\pi}^2$ of the np CEX Uppsala data.}
\label{fig:Ashmore}
\end{center}
\end{figure}

A second method, which should improve the convergence of the Chew
extrapolation, is  the Ashmore method~\cite{Ashmore62}. It parameterises 
${\rm d}\sigma/{\rm d}\Omega(x)$ in terms of a pion Born amplitude with
the addition of a background term. Here we use the 
regularised pion Born amplitudes as   
given in Ref.~\cite{Gibbs94}. One 
expects also an important contribution to the np CEX
from the $\rho$-meson exchange. We then use for the Ashmore background 
amplitude a pole term  with adjustable 
strength simulating this $\rho$-meson exchange. This expression is 
fitted to the data and gives in principle a 
model-independent result for the coupling constant. More 
physics is built into the procedure, so fewer terms should be 
needed. More detailed expressions can be found in Ref~\cite{Rahm98}.

In order to obtain an 
improvement in the extrapolation  we have introduced
the Difference Method~\cite{Eri95}. It is based 
on the Chew function, but it uses the fact that an important part of the
cross  section behaviour is described by models with exactly known values for 
the coupling constant. It applies the Chew method to the 
{\it difference} between the function $y(x)$ of a model and that
of the experimental data, i.e., 
\begin{equation}
y_{Model}(x)-y_{Exp}(x) = \sum_{{\rm i=0}}^{{\rm n-1}} d_ix^i.
\label{eq:Difference}
\end{equation}
If $g_R$ of Eq.~(\ref{eq:Chew}) is replaced by the model value $g_{Model}$,
one has at the pion pole,
\begin{equation}
y_{Model}(0)-y_{Exp}(0) \equiv d_0 \equiv 
\frac{g_{Model}^4-g^4_{\pi^\pm}}{g_{Model}^4}.
\label{eq:polediff}
\end{equation}
This  should decrease systematic extrapolation uncertainties and remove a 
substantial part of the non-OPE information at large  momentum 
transfers. We have formally not 
introduced a model dependence by using such a comparison function and 
such procedures are used in many contexts of physics to 
obtain better transparency and precision. One has to 
calibrate the method, that is, to find the precision to which the 
coupling constant can be determined and the possible systematic uncertainties
that are associated with the extrapolation procedure.

\subsection{Application to models}
\label{models}

\begin{table}[ht]
\begin{center}
\caption{Coupling constants obtained from $n$-term polynomial fits 
to the 31 points of several `pseudo-data' models at 162~MeV for the reduced 
range $0<q^2<4\ m_{\pi}^2$ with the three extrapolation methods. 
These models are $NN$ energy dependent PWA's of 
R. A. Arndt~\protect\cite{Arndt99b}. The unknown coupling
constant of the model `DA99' is to be determined while the reference models
A13.75 and A12.83g
 have $g^2_{\pi^\pm}/4\pi$ of 13.75 and 12.83 respectively. The
column $\delta g^2_{\pi^\pm}$ is the systematic shift from the
true model value. In boldface the value at $\chi^2/N_{df}$ = 1.00, retained
for the determination of $g^2_{\pi^\pm}/4\pi$.}

\vspace{3mm}
\small
\begin{tabular}{c c c c c c c c c }
\hline
\hline
 n &$\chi^2/N_{df}$ & \multicolumn{1}{c}{$g^2_{\pi^\pm}/4\pi$}
   & $\chi^2/N_{df}$ & \multicolumn{1}{c}{$g^2_{\pi^\pm}/4\pi$}
   & $\delta g^2_{\pi^\pm}$  
   & $\chi^2/N_{df}$ & \multicolumn{1}{c}{$g^2_{\pi^\pm}/4\pi$}
   & $\delta g^2_{\pi^\pm}$\\  
   & \multicolumn{2}{c}{`DA99'}
   & \multicolumn{3}{c}{`A12.83g'}
   & \multicolumn{3}{c}{`A13.75'} \\
  \multicolumn{9}{c}{Chew Method}\\
\hline
\ignorespaces
 4 & 1.18  &      $11.76 \pm 0.38$
   & 1.13  &      $10.40 \pm 0.43$
   & 2.43
   & 1.16  &      $11.27 \pm 0.40$
   & 2.48\\
 5 & 1.00  &      $13.56 \pm 0.84$
   & 1.00  &      $12.12 \pm 0.94$
   & 0.71
   & 1.00  &      $13.04 \pm 0.87$
   & 0.71\\
\\   
 \multicolumn{9}{c}{Ashmore Method} \\
\hline
 4 & 1.01  &      $13.46 \pm 0.47$
   & 1.01  &      $12.00 \pm 0.54$
   & 0.83
   & 1.01  &      $13.01 \pm 0.50$
   & 0.74\\
 5 & 1.00  &      $13.89 \pm 0.94$

   & 1.00  &      $12.48 \pm 1.11$
   & 0.35
   & 1.00  &      $13.40 \pm 1.03$
   & 0.35\\
\\   
  \multicolumn{9}{c}{Difference Method}\\
   & \multicolumn{2}{c}{A12.83g $-$ `DA99'}  
   & \multicolumn{2}{c}{A13.75 $-$ `DA99'}
   &
   & \multicolumn{3}{c}{A13.75 $-$ `A12.83g'}\\
\hline
 2 & 3.32  &      $12.15 \pm 0.06$
   & 1.34  &      $13.51 \pm 0.06$
   &
   & 1.88  &      $14.15 \pm 0.05$
   & 1.32\\
 3 & 1.30  &      $13.09 \pm 0.13$
   & 1.04  &      $13.85 \pm 0.12$
   &
   & 1.12  &      $13.62 \pm 0.13$
   & 0.79\\
 4 & 1.00  &   {\bf 13.97}   $ \pm $ {\bf 0.32}
   & 1.00  &     {\bf 14.16} $ \pm$  {\bf 0.32}
   &
   & 1.00  &      $13.05 \pm 0.34$
   & 0.22\\
\hline
\hline
\end{tabular}
\normalsize
\end{center}
\protect\label{tab:difdared}
\end{table}

We now want to cross-check to which precision can work the difference method
in the extrapolation to the pion pole. We shall first 
apply it to models where the coupling is exactly
known. This allows to investigate its properties and in particular its
systematics.
In order to determine the systematic 
uncertainties in the procedures we have generated pseudo-data with 
uncertainties corresponding to the Uppsala 162 MeV experiment from 10000 
computer simulations using exact data points from different models
 with a Gaussian, random error 
distribution~\cite{Gibbs94a}. One has for a given pseudo-measurement m,
\begin{equation}
y^{Pseudo-data}_{m}(x) = y^{Model}(x)+\Delta y_m(x)
\label{pseudodata}
\end{equation}
with
\begin{equation}
\Delta y_m(x)=\Delta y^{Uppsala}(x) \sqrt{-2 Log ( R_{1m})} cos(\pi R_{2m}).
\label{deltay}
\end{equation}
In Eq.~\ref{deltay} $R_{im}$, for i=1, 2, are random
numbers between 0 and 1  when m varies from 1 to 10000.

In the context of the workshop we have asked R. A. Arndt to provide us with
different models from
his $NN$ energy dependent PWA. One of the models, which we call A13.75,
corresponds to the energy-dependent PWA of the $pp$ and np data from 0 to 400
MeV~\cite{Arndt99b} with a minimisation on  $g^2_{\pi^\pm}/4\pi$~\cite{Arn99}.
 The minimum $\chi^2 $ on the $NN$ data is obtained for a 
 coupling constant of 13.75. The second model we consider is
 built from the
previous one  with all parameters kept fixed,
except the value of the $\pi NN$ coupling constant which is lowered down
from 13.75 to 12.83, let us denote this model as A12.83g. A third
model, denoted DA99, was also given to us but with an unknown 
coupling constant.
We shall here apply the three methods described 
in section~\ref{extrap_meth} in order to attempt
to determine this coupling constant,
considering the predictions of this model as pseudo-data. The differential
cross section at 162 MeV of the model DA99 is
compared to that of the Uppsala experiment in Fig.~\ref{fig:162da99}.  

\begin{table}[ht]
\begin{center}
\caption{Same as for Table~2 
but for the 54 `pseudo-data' of
the full range $0<q^2<10.1\ m_{\pi}^2$.}
\protect\label{tab:difdafull}
\vspace{3mm}

\small
\begin{tabular}{c c c c c c c c c }
\hline
\hline
 n &$\chi^2/N_{df}$ & \multicolumn{1}{c}{$g^2_{\pi^\pm}/4\pi$}
   & $\chi^2/N_{df}$ & \multicolumn{1}{c}{$g^2_{\pi^\pm}/4\pi$}
   & $\delta g^2_{\pi^\pm}$  
   & $\chi^2/N_{df}$ & \multicolumn{1}{c}{$g^2_{\pi^\pm}/4\pi$}
   & $\delta g^2_{\pi^\pm}$\\  
   & \multicolumn{2}{c}{`DA99'}
   & \multicolumn{3}{c}{`A12.83g'}
   & \multicolumn{3}{c}{`A13.75'} \\
  \multicolumn{9}{c}{Chew Method}\\
\hline
\ignorespaces
 5 & 1.13  &      $12.18 \pm 0.29$
   & 1.11  &      $10.74 \pm 0.33$
   & 2.09
   & 1.12  &      $11.65 \pm 0.30$
   & 2.10\\
 6 & 1.02  &      $13.23 \pm 0.51$
   & 1.01  &      $11.80 \pm 0.57$
   & 1.03
   & 1.02  &      $12.71 \pm 0.53$
   & 1.04\\
 7 & 1.00  &      $13.86 \pm 0.87$
   & 1.00  &      $12.42 \pm 0.97$
   & 0.41
   & 1.00  &      $13.34 \pm 0.90$
   & 0.41\\
\\   
 \multicolumn{9}{c}{Ashmore Method} \\
\hline
 4 & 3.38  &      $11.96 \pm 0.33$
   & 4.13  &      $\ 9.62 \pm 0.41$
   & 3.21
   & 3.65  &      $11.13 \pm 0.35$
   & 2.62\\
 5 & 1.05  &      $14.25 \pm 0.33$
   & 1.03  &      $12.64 \pm 0.90$
   & 0.19
   & 1.04  &      $13.69 \pm 0.35$
   & 0.06\\
 6 & 1.03  &      $13.82 \pm 0.78$
   & 1.02  &      $12.03 \pm 0.38$
   & 0.80
   & 1.02  &      $12.91 \pm 0.84$
   & 0.84\\
\\   
  \multicolumn{9}{c}{Difference Method}\\
   & \multicolumn{2}{c}{A12.83g $-$ `DA99'}  
   & \multicolumn{2}{c}{A13.75 $-$ `DA99'}
   &
   & \multicolumn{3}{c}{A13.75 $-$ `A12.83g'}\\
\hline
 3 & 2.68  &      $12.23 \pm 0.09$
   & 1.24  &      $13.55 \pm 0.08$
   &
   & 1.65  &      $14.14 \pm 0.08$
   & 1.28\\
 4 & 1.26  &      $13.31 \pm 0.15$
   & 1.04  &      $13.93 \pm 0.14$
   &
   & 1.10  &      $13.48 \pm 0.15$
   & 0.65\\
 5 & 1.00  &   {\bf 14.07} $\pm $ {\bf 0.25}
   & 1.00  &    {\bf  14.20} $\pm $ {\bf 0.25}
   &
   & 1.00  &      $12.99 \pm 0.27$
   & 0.16\\
\hline
\hline
\end{tabular}
\normalsize
\end{center}
\end{table}

Results  are listed  in Tables~2 
and~3, 
$n$ being the number of terms in the polynomial 
fit of the 'pseudo-data'. As for each calculation we performed 10000 
pseudo-experiments,
$\chi^2/N_{df}$ 
is the average $\chi^2$ per degrees of 
freedom, $g^2_{\pi^\pm}/4\pi$ is the mean value of the coupling constant 
and the errors quoted are the standard 
deviations which, in fact, are very close to the average value of the 
error of every pseudo-experiment.  We also 
give, for the models A12.83g and A13.75,
the systematic deviation $\delta g^2_{\pi^\pm}$ of the mean value 
from the true value in the model. We have then a control
on systematic extrapolation errors and can calibrate the corresponding 
corrections. The data is grouped in two
intervals, the first one with $0 < q^2 < 4\ m_{\pi}^2$  called 
'reduced range'  with 31 data 
points, corresponding to the previous Uppsala experiment~\cite{Eri95} and
 the second one with
$0 < q^2 < 10.1\ m_{\pi}^2$ denoted 'full range' with 54 data points
corresponding to  the latest experiment~\cite{Rahm98,Olsson99}. 
This allows to examine the sensitivity and stability of the 
extrapolation to a given cut in momentum transfer and to check
that it is the small $q^2$ region that carries an important part
 of the pion pole 
information. As a function of $n$ the behaviour of $\chi^2/N_{df}$  is 
characteristic: it drops quickly with increasing $n$ to a value close 
to unity. Additional terms give only small benefits, and the data become 
over-parameterised. One can then adopt different statistical 
strategies leading to similar results. One is to take 
results at the minimum $\chi^2/N_{df}$. This minimum is usually a 
shallow one, and  values of $n$ close to 
$n$ of $\chi^2/N_{df}$ minimum are almost equally probable statistically. 
Another possibility is to pick up $g^2_{\pi^\pm}/4\pi$ from one of 
the smallest 
values of $n$ consistent with a $\chi^2/N_{df}$ well within the range 
expected from the experimental sample. We recall the reader that here 
there is about 47\% probability of the experimental 
$\chi^2/N_{df}$ to be larger than unity, and about 25\% for it to be 
larger than 1.15.

\begin{figure}[htb]
\begin{center}
\includegraphics
[angle=90,width=14cm]{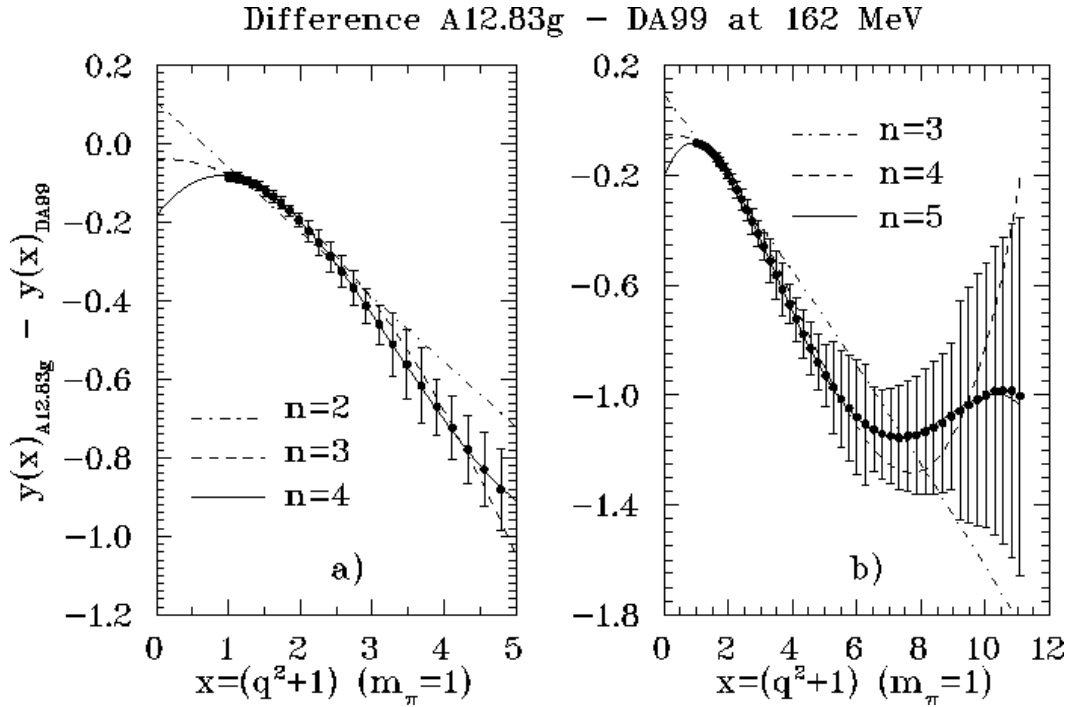}
\caption{Extrapolations of the Chew function $y(q^2)$ to the pion pole 
at 162 MeV with the Difference Method on the model
DA99~\protect\cite{Arndt99b} for $n$-term
polynomial fits and for the reduced, a), $0 < q^2 < 4\ m_{\pi}^2$ and
 full, b),  $0 < q^2 < 10.1\ m_{\pi}^2$ ranges. The
reference function is the PWA A12.83g~\protect\cite{Arndt99b}.
Here exact prediction of the model is used, the error at each point being
that of the corresponding 162 MeV Uppsala data 
point~~\protect\cite{Rahm98,Olsson99}.}
\label{fig:12gda}
\end{center}
\end{figure}

For the Chew method a good fit is performed with a fourth order 
polynomial in $q^2$ for the reduced range, but with a large systematic 
downward shift of 0.71 for both 'A12.83g' and 'A13.75' PWA's  as compared
to the original model values. With a third order polynomial  fit the
statistical error becomes smaller, but the systematic shift is 
unreasonably large viz. 2.43 and 2.48 respectively. In the full  range ,
one or two more terms are needed to obtain a  good fit. In any case a
systematic shift remains even when a perfect  fit is obtained, but at the
minimum $\chi^2$ it is always less than the statistical and extrapolation 
uncertainty. The 'DA99' pseudo-data for both ranges give slightly different
results for $g^2_{\pi^\pm}/4\pi$ at minimum $\chi^2$, but if one applies
the corresponding systematic shift one obtains the same  value of
14.27(86).The  statistical and  extrapolation error is rather large so
we do not obtain a precise determination of the coupling constant using this
method. Fig.~\ref{fig:Chew} shows the fit of the model DA99
in the reduced range together
with its  Chew-function
(Eq.~\ref{eq:Chew}) extrapolations 
for n=4 (dotted line) and n=5 (solid line). The n=5 fit
is better for $q^2$ above 2 m$_\pi^2$.

\begin{figure}[htb]
\begin{center}
\includegraphics
[angle=90,width=14cm]{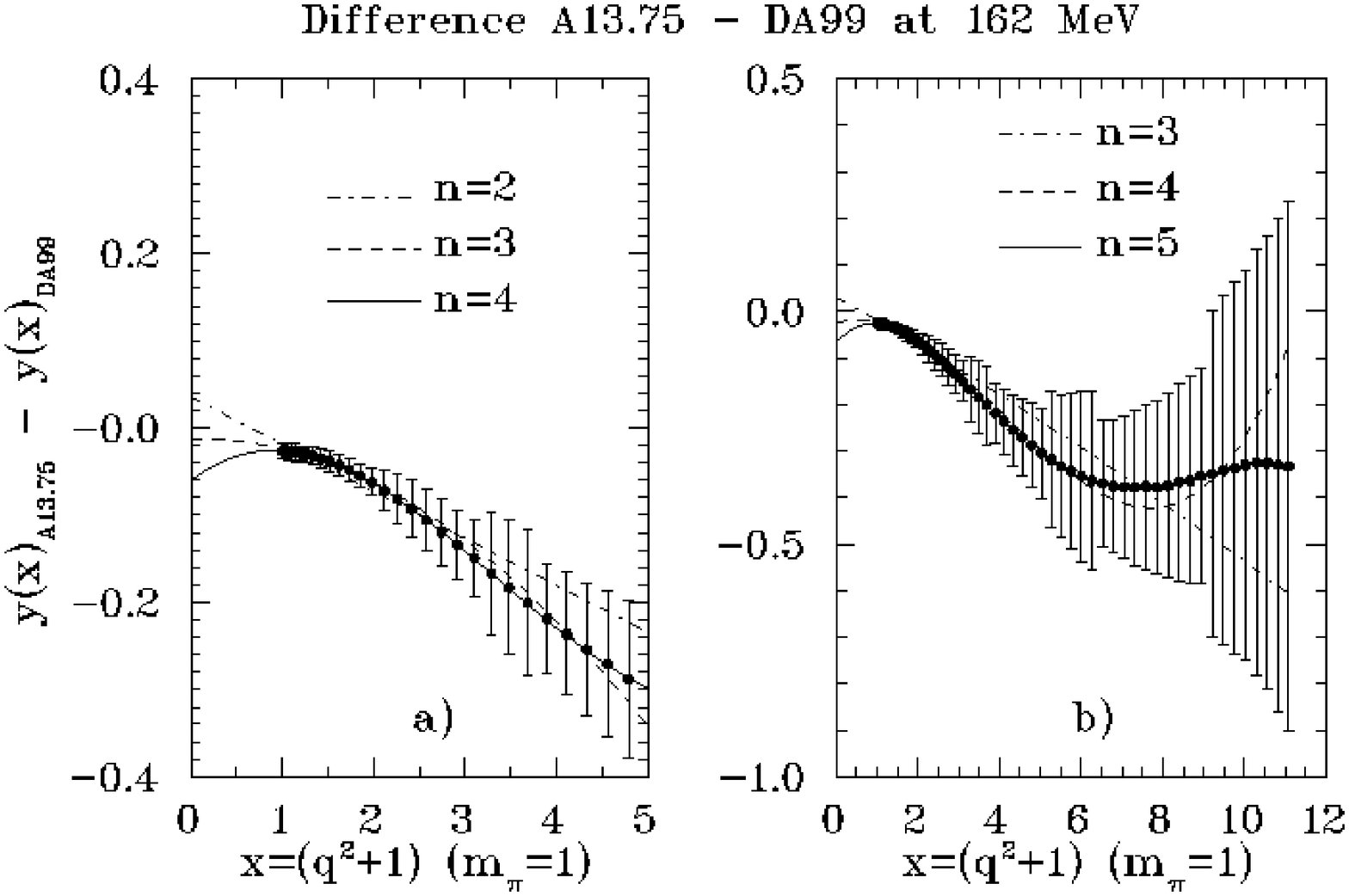}
\caption{As in Fig.~\protect\ref{fig:12gda}
but for the reference model 'A13.75'~\protect\cite{Arndt99b}.}
\label{fig:13da}
\end{center}
\end{figure}

For the Ashmore method a good description is achieved in the
reduced range with one term less in the expansion. This shows that the
physics beyond the $\pi$-exchange is reasonably described by the
$\rho$-exchange, as anticipated in section~\ref{extrap_meth}.
The systematic shifts are similar to
those  of the Chew method. The
statistical and extrapolation error,
when one stands close to the minimum $\chi^2$,
is however smaller. Once corrected by
the mean value of the systematic shifts of the models the
pseudo-data gives a $g^2_{\pi^\pm}$ of 14.25(47) with a central
value very close to the previous result. In the full range the needed number
of  terms to get a good $\chi^2$
is also smaller and the corrected value for n=5 is 14.37(35).
Although the statistical and extrapolation
accuracy has improved, this method  also appears to lack the high accuracy
we would like to have. The excellent fit to the model
 DA99 is drawn in Fig.~\ref{fig:Ashmore} for the full range with the Ashmore
parameterisation~\cite{Rahm98} with 5 terms.

The Difference Method should need less terms in the polynomial 
expansion than  the two above methods, and this will give a smaller,
statistical  extrapolation error. Recall that {\it the statistical
extrapolation errors are  
only meaningful if $\chi^2/N_{df}$ is close to 1}. The fact that
the angular distributions from the pseudo-data  and models might be alike
can help, in particular for large $q^2$. This can add 
more physical information without introducing in principle any 
model dependence. We use  the two reference models 
studied above.  Results are  given in Tables~2 
and~3 
for the  reduced and full
ranges, respectively.
The pion-pole extrapolations 
of the $n$-term polynomial fits
of the Difference Method, on the
the reduced and full  ranges for the model DA99,
are shown in Figs.~\ref{fig:12gda} 
and~\ref{fig:13da}  for the comparison models A12.83g and  A13.75,
respectively.
 The error bars increase  at large $x$, which is due to
the multiplication of the cross  section by $x^2$, giving a smaller
weight for the large $q^2$  region when extrapolating. The difference
behaviour is slightly smoother with the reference model 'A13.75', however
it can be seen that in all cases 4 and 5 terms are necessary to have a good
fit in the reduced and full range, respectively. In the full range, as one
has more points,  one gets a better statistical extrapolation error, this
applies also to the difference between reference models which gives a 
 check on the systematic uncertainties in the extrapolation. Both ranges
have systematic shifts below 1\%.  

 Averaging the
values at minimum $\chi^2 $ (values in boldface in Tables 2 and 3) 
from the Difference-Method extrapolations  one gets, over the reduced range, 
\begin{eqnarray*}
\sqrt{N} g^2_{\pi^\pm}/4\pi & = & 14.07 \pm 0.32\ (\mbox{statistical +
extrapolation})  \\
         & & \pm 0.11\ (\mbox{systematic}) \pm 0.16\ (\mbox{normalisation})
	  \\
                            & = & 14.07 \pm 0.37,
			    \end{eqnarray*}
i.e. an accuracy of 2.6 \%,
and over the full range,
\begin{eqnarray*}
\sqrt{N} g^2_{\pi^\pm}/4\pi
& = &14.14\pm 0.25\ (\mbox{stat. + extr.}) \pm 0.08\ (\mbox{syst.})
                                \pm 0.16\ (\mbox{norm.})  \\
& = &14.14 \pm 0.31,
\end{eqnarray*}
which corresponds to an accuracy of 2.2 \%. The results are fairly
close which substantiates our statement on the relevant information
being nearly  entirely at low $q^2$. In view of the somewhat larger
extrapolation  uncertainty in the case of the reduced range, we take the
 {\it full  range value}, {\bf 14.14(31)}, which compares quite well with the
{\it  exact value of the model} DA99 which is {\bf 14.28}. Using the 
experimentally given  statistics of the 162 MeV Uppsala data, the
difference method has allowed us to determine the unknown DA99 coupling to
less than 1 \% within an uncertainty of 2.2 \%.

\subsection{Application to data}
\label{data}

\begin{table}[htb]
\caption{$g^{2}_{\pi^\pm}/4\pi$ and $f^{2}_{c}/4\pi$ of the PWA comparison 
	models~\protect\cite{Arndt99b} used 
here together with their total $\chi^{2}$ and $\chi^{2}$/data on the 
3747 np data below 400 MeV.}
\protect\label{tab:Arndtpwa}
\vspace{3mm}
\begin{center}
\begin{tabular}{c|c c c c c}
	\hline
	\hline
	Model & A12.83g & A12.83 & A13.75 & A14.28 & A14.28g  \\
	\hline
	$g^{2}_{\pi^\pm}/4\pi$ & 12.83 & 12.83 & 13.75 & 14.28 & 14.28  \\
	$f^{2}_{c}/4\pi$ & 0.071 & 0.071 & 0.076 & 0.079 & 0.079  \\
	$\chi^{2}_{np}(total)$ & 9282 & 5087 & 4975 & 4958 & 6149  \\
	$\chi^{2}_{np}/data$ & 2.48 & 1.36 & 1.33 & 1.32 & 1.64  \\
	\hline
	\hline
\end{tabular}
\end{center}
\end{table}

\begin{figure}[htb]
\begin{center}
\includegraphics
[angle=90,width=14cm]{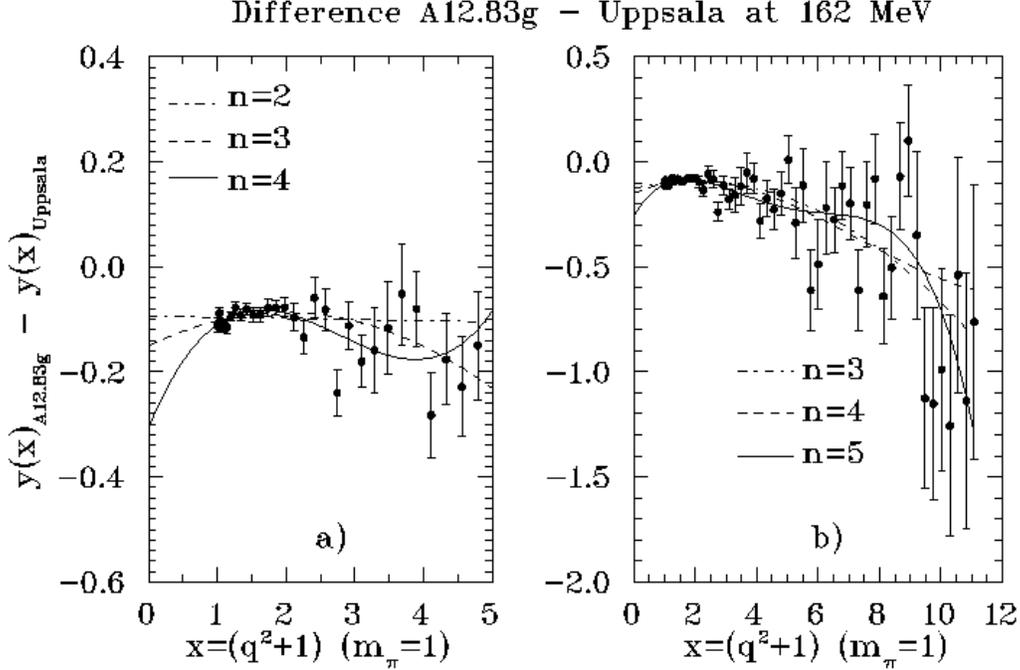}
\caption{As in Fig. \protect\ref{fig:12gda} but for the difference
between model A12.83g~\protect\cite{Arndt99b} and Uppsala 
data~\protect\cite{Rahm98,Olsson99}.}
\label{fig:12.83g}
\end{center}
\end{figure}

We shall now apply the difference method to the 162 MeV Uppsala 
data~\cite{Rahm98,Olsson99} to cross-check the precision of the 
$g^{2}_{\pi^\pm}/4\pi$ determination of Ref.~\cite{Rahm98}. We here consider
different reference models than those used in~\cite{Rahm98}. Besides 
the three comparison models considered in the previous section we 
also use two more PWA fit of the $pp$ and np data from 0 to 400 MeV 
with $g^{2}_{\pi^\pm}/4\pi$ fixed to 12.83 (model A12.83) and 14.28 
(model A14.28)~\cite{Arndt99b}. In Table~4 
we give the total $\chi^{2}$ and the $\chi^{2}$/data on the 3747 np data 
considered for all the models we study here. We also recall their 
$g^{2}_{\pi^\pm}/4\pi$ together with the corresponding pseudovector 
coupling $f^{2}_{c}/4\pi$, related to $g^{2}_{4\pi^\pm}/4\pi$ by
$g^{2}_{\pi^\pm}=f^{2}_{c}(2M_{M_{N}}/m_{\pi^\pm})^{2}$ with 
$M_{N}$ average proton-neutron mass. 
As for the model A12.83g, the model A14.28g 
(DA99) was obtained from the model A13.75 with all parameters kept 
fixed but increasing the $\pi NN$ coupling from 13.75 to 14.28.
Predictions of all these comparison models for the 162 MeV Uppsala data
are shown in Fig.~\ref{fig:162da99}.

\begin{table}[ht]
\begin{center}
\caption{As in Table~2, 
but, i)  for the difference method
only, ii) the 31 points of the pseudo-data of the model DA99
are replaced by those of the 162 MeV Uppsala
data~\protect\cite{Rahm98,Olsson99}, iii) with 3 more Arndt PWA reference
models~\protect\cite{Arndt99b},
viz., A12.83, A14.28 and A14.28g (same as `DA99'). Models A12.83 and
 A14.28 (as also A14.28g)
 have $g^2_{\pi^\pm}/4\pi$ of 12.83 and 14.28, respectively.
 In boldface the experimental values retained for the determination of
$g^2_{\pi^\pm}/4\pi$.} 
\protect\label{tab:difupred}

\vspace{3mm}

\small
\begin{tabular}{c c c c c c c c c }
\hline
\hline
  \multicolumn{9}{c}{Difference Method}\\
 n &$\chi^2/N_{df}$ & \multicolumn{1}{c}{$g^2_{\pi^\pm}/4\pi$}
   & $\chi^2/N_{df}$ & \multicolumn{1}{c}{$g^2_{\pi^\pm}/4\pi$}
   & $\delta g^2_{\pi^\pm}$  
   & $\chi^2/N_{df}$ & \multicolumn{1}{c}{$g^2_{\pi^\pm}/4\pi$}
   & $\delta g^2_{\pi^\pm}$\\  
\hline
\ignorespaces
   & \multicolumn{2}{c}{A12.83g $-$ Uppsala}  
   & \multicolumn{2}{c}{A12.83 $-$ Uppsala}
   &
   & \multicolumn{2}{c}{A13.75 $-$ Uppsala}
   &\\
\hline
 2 & 1.60  &      $13.45 \pm 0.05$
   & 1.13  &      $13.85 \pm 0.05$
   &
   & 1.13  &      $14.69 \pm 0.05$
   & \\
 3 & 1.34  &      $13.79 \pm 0.13$
   & 1.15  &      $13.75 \pm 0.13$
   &
   & 1.06  & {\bf 14.50} $\pm$ {\bf 0.12}
   & \\
 4 & 1.03  & {\bf 14.69} $\pm$ {\bf 0.31}
   & 1.03  & {\bf 14.37} $\pm$ {\bf 0.31}
   &
   & 1.03  &      $14.87 \pm 0.30$
   & \\
 5 & 1.07  &      $14.50 \pm 0.79$
   & 1.07  &      $14.30 \pm 0.80$
   &
   & 1.06  &      $14.54 \pm 0.78$
   & \\
\hline
\ignorespaces
   & \multicolumn{2}{c}{A14.28 $-$ Uppsala}  
   & \multicolumn{2}{c}{A14.28g $-$ Uppsala}
   &
   & \multicolumn{3}{c}{A13.75 $-$ `12.83'}\\
\hline
 2 & 1.17  &      $15.20 \pm 0.05$
   & 1.72  &      $15.42 \pm 0.05$
   &
   & 1.05  &      $13.76 \pm 0.05$
   & 0.93\\
 3 & 1.02  & {\bf 14.96} $\pm$ {\bf 0.12}
   & 1.00 &  {\bf 14.94} $\pm$ {\bf 0.12}
   & 
   & 1.03  &      $13.65 \pm 0.13$
   & 0.82\\
 4 & 1.04  &      $15.19 \pm 0.30$
   & 1.03  &      $15.00 \pm 0.30$
   &
   & 1.01  &      $13.40 \pm 0.33$
   & 0.57\\
 5 & 1.06  &      $14.72 \pm 0.77$
   & 1.06  &      $14.59 \pm 0.78$
   &
   & 1.00  &      $13.09 \pm 0.87$
   & 0.26\\
\hline
\ignorespaces
   &
   &   
   & \multicolumn{3}{c}{A12.83 $-$ `A14.28'}
   & \multicolumn{3}{c}{A13.75 $-$ `A14.28'}\\
\hline
 2 &       &       
   & 1.14  &      $12.86 \pm 0.06$    
   &  -1.42
   & 1.02  &      $13.76 \pm 0.05$
   & -0.52\\
 3 &       &      
   & 1.07  &      $13.04 \pm 0.13$
   & -1.24
   & 1.01  &      $13.83 \pm 0.12$
   & -0.45\\
 4 &       &
   & 1.01  &      $13.43 \pm 0.33$
   & -0.85
   & 1.00  &      $13.96 \pm 0.32$
   & -0.32\\
 5 &       &
   & 1.00  &      $13.83 \pm 0.83$
   & -0.45
   & 1.00  &      $14.09 \pm 0.81$
   & -0.19\\
\hline
\hline
\end{tabular}
\normalsize
\end{center}
\end{table}

Results for the reduced and full range are listed in
Tables~5 
and~6 
respectively.
The A12.83g comparison model requires 4 terms in the reduced range and
5 in the full one. Figure~\ref{fig:12.83g} shows that there is some
edge effect in the reduced range at $q^{2}=4m^{2}_{\pi}$.
For the model A12.83 one needs also 4 terms in the reduced range. The
obtained value, 14.37(31) is quite consistent with those determined in the
model A12.83g, either in the reduced range, 14.69(31) or in the full range,
14.41(24).
In the
full range the minimum is very shallow  and the $n=4\
(\chi^{2}/N_{df}=1.18)$ and $n=5\ (\chi^{2}/N_{df}=1.16)$ values,
13.65(14) and 13.94(25), respectively, are compatible.
The lower value, 13.65(14), is however not compatible with
the value, 14.37(31) of the reduced range at $\chi^2/N_{df}$ minimum,
We shall then retain, in the full range for the A12.83 model, 
the $n=5$ determination, 13.94(25). 
Figure~\ref{fig:12.83} shows as before the necessity to have at least 4 
or 5 terms to obtain a good fit. In the reduced range 3 terms are here 
sufficient with the reference model A13.75 leading to 
$g^{2}_{\pi^\pm}/4\pi=14.50(12)$ while in the full range $\chi^{2}$ 
minimum is reached with 4 terms with a value of 14.38(14). 
Figure~\ref{fig:13.75} shows in both range a smoother behaviour than 
before. In the A14.28 and A14.28g case one needs the same number of 
terms as for A13.75 but the corresponding $g^{2}_{\pi^\pm}/4\pi$ are 
somewhat larger and not always compatible with previous values.
Curves for the extrapolation are shown in Figs.~\ref{fig:14.28}
and~\ref{fig:14.28g}. Applying the difference method between models (see 
also Tables~2 
and~3) 
shows relatively large systematics which allows to understand this dispersion.

\begin{table}[htb]
\begin{center}
\caption{As in Table~5 
but for the 54  Uppsala data points at 162 Mev of the full
range $0<q^2<10.1\ m_{\pi}^2$.} 

\vspace{3mm}

\small
\begin{tabular}{c c c c c c c c c }
\hline
\hline
  \multicolumn{9}{c}{Difference Method}\\
 n &$\chi^2/N_{df}$ & \multicolumn{1}{c}{$g^2_{\pi^\pm}/4\pi$}
   & $\chi^2/N_{df}$ & \multicolumn{1}{c}{$g^2_{\pi^\pm}/4\pi$}
   & $\delta g^2_{\pi^\pm}$  
   & $\chi^2/N_{df}$ & \multicolumn{1}{c}{$g^2_{\pi^\pm}/4\pi$}
   & $\delta g^2_{\pi^\pm}$\\  
\hline
\ignorespaces
   & \multicolumn{2}{c}{A12.83g $-$ Uppsala}  
   & \multicolumn{2}{c}{A12.83 $-$ Uppsala}
   &
   & \multicolumn{2}{c}{A13.75 $-$ Uppsala}
   & \\
\hline
 3 & 1.30  &      $13.65 \pm 0.08$
   & 1.36  &      $14.02 \pm 0.08$
   &
   & 1.43  &      $14.84 \pm 0.07$
   & \\
 4 & 1.30  &      $13.79 \pm 0.14$
   & 1.18  &      $13.65 \pm 0.14$
   &
   & 1.12  & {\bf 14.38} $\pm$ {\bf 0.14}
   & \\
 5 & 1.14  & {\bf 14.41} $\pm$ {\bf0.24}
   & 1.16  & {\bf 13.94} $\pm$ {\bf0.25}
   &
   & 1.14  &      $14.53 \pm 0.24$
   & \\
 6 & 1.15  &      $14.74 \pm 0.46$
   & 1.16  &      $14.42 \pm 0.47$
   &
   & 1.15  &      $14.80 \pm 0.45$
   & \\
\hline
   & \multicolumn{2}{c}{A14.28 $-$ Uppsala}  
   & \multicolumn{2}{c}{A14.28g $-$ Uppsala}
   &
   & \multicolumn{3}{c}{A13.75 $-$ `12.83'}\\
\hline
 3 & 1.51  &      $15.33 \pm  0.07$
   & 2.15  &      $15.53 \pm  0.07$
   &
   & 1.03  &      $13.74 \pm 0.08$
   & 0.91\\
 4 & 1.11  & {\bf 14.83} $\pm$ {\bf 0.13}
   & 1.12  & {\bf 14.75} $\pm$ {\bf 0.13}
   &
   & 1.02  &      $13.63 \pm 0.14$
   & 0.80\\
 5 & 1.13  &      $14.89 \pm 0.24$
   & 1.13  &      $14.62 \pm 0.24$
   &
   & 1.01  &      $13.49 \pm 0.26$
   & 0.66\\
 6 & 1.15  &      $15.05 \pm 0.45$
   & 1.15  &      $14.86 \pm 0.45$
   &
   & 1.00  &      $13.27 \pm 0.50$
   & 0.44\\
\hline
   &   
   & 
   &\multicolumn{3}{c}{A12.83 $-$ `A14.28'}
   & \multicolumn{3}{c}{A13.75 $-$ `A14.28'}\\
\hline
 3 &       &       
   & 1.08  &      $12.89 \pm 0.09$    
   &  -1.41
   & 1.01  &      $13.77 \pm 0.08$
   & -0.51\\
 4 &       &      
   & 1.04  &      $13.07 \pm 0.15$
   & -1.21
   & 1.01  &      $13.83 \pm 0.14$
   & -0.45\\
 5 &       &
   & 1.02  &      $13.30 \pm 0.26$
   & -1.13
   & 1.00  &      $13.92 \pm 0.25$
   & -0.36\\
 6 &       &
   & 1.01  &      $13.62 \pm 0.49$
   & -0.66
   & 1.00  &      $14.02 \pm 0.48$
   & -0.26\\
\hline
\hline
\end{tabular}
\normalsize
\end{center}
\protect\label{tab:difupfull}
\end{table}

\begin{figure}[htb]
\begin{center}
\includegraphics
[angle=90,width=14cm]{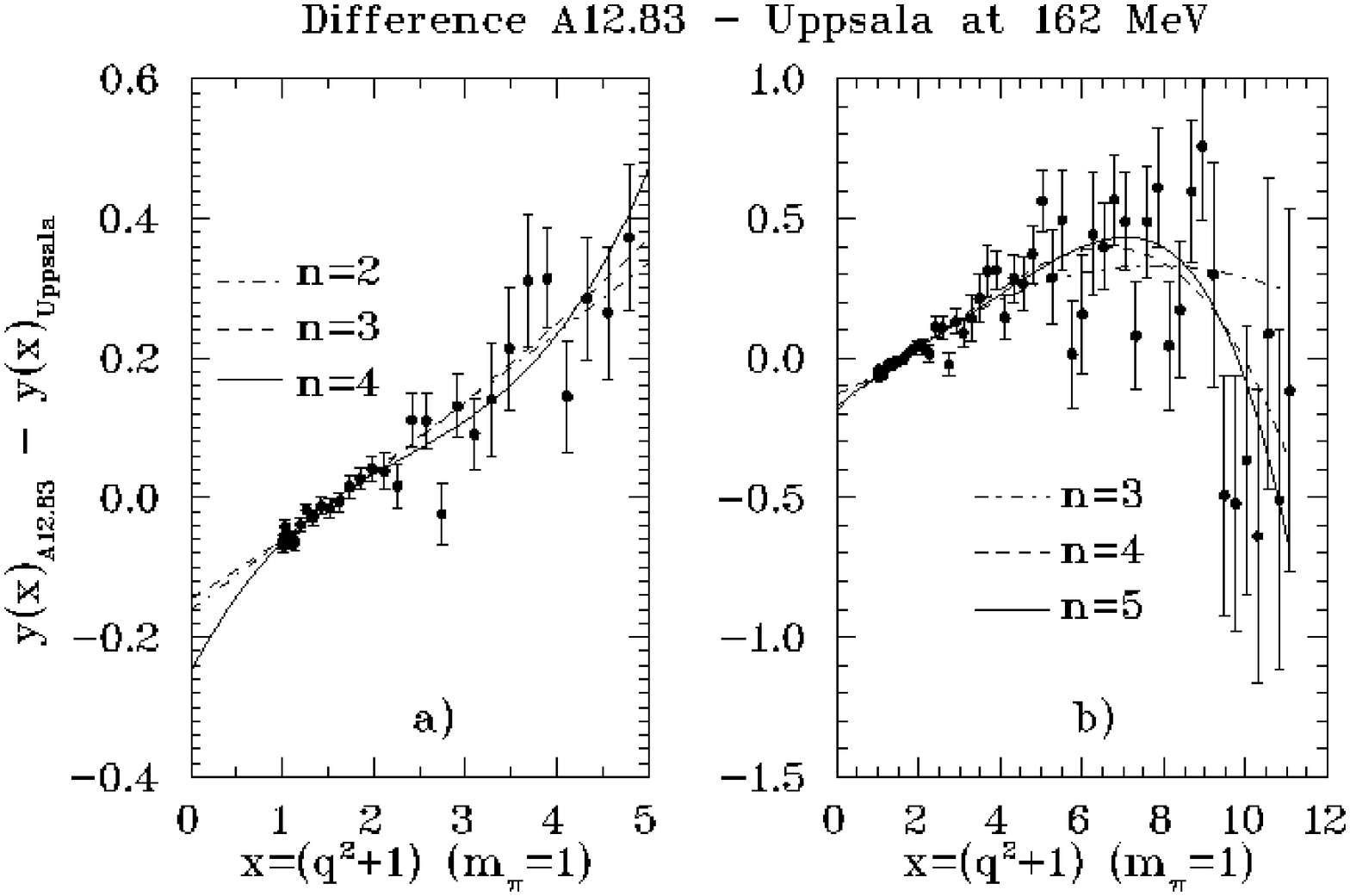}
\caption{As in Fig. \protect\ref{fig:12gda} but for the difference
between model A12.83~\protect\cite{Arndt99b} and Uppsala 
data~\protect\cite{Rahm98,Olsson99}.}
\label{fig:12.83}
\end{center}
\end{figure}

\begin{figure}[htb]
\begin{center}
\includegraphics
[angle=90,width=14cm]{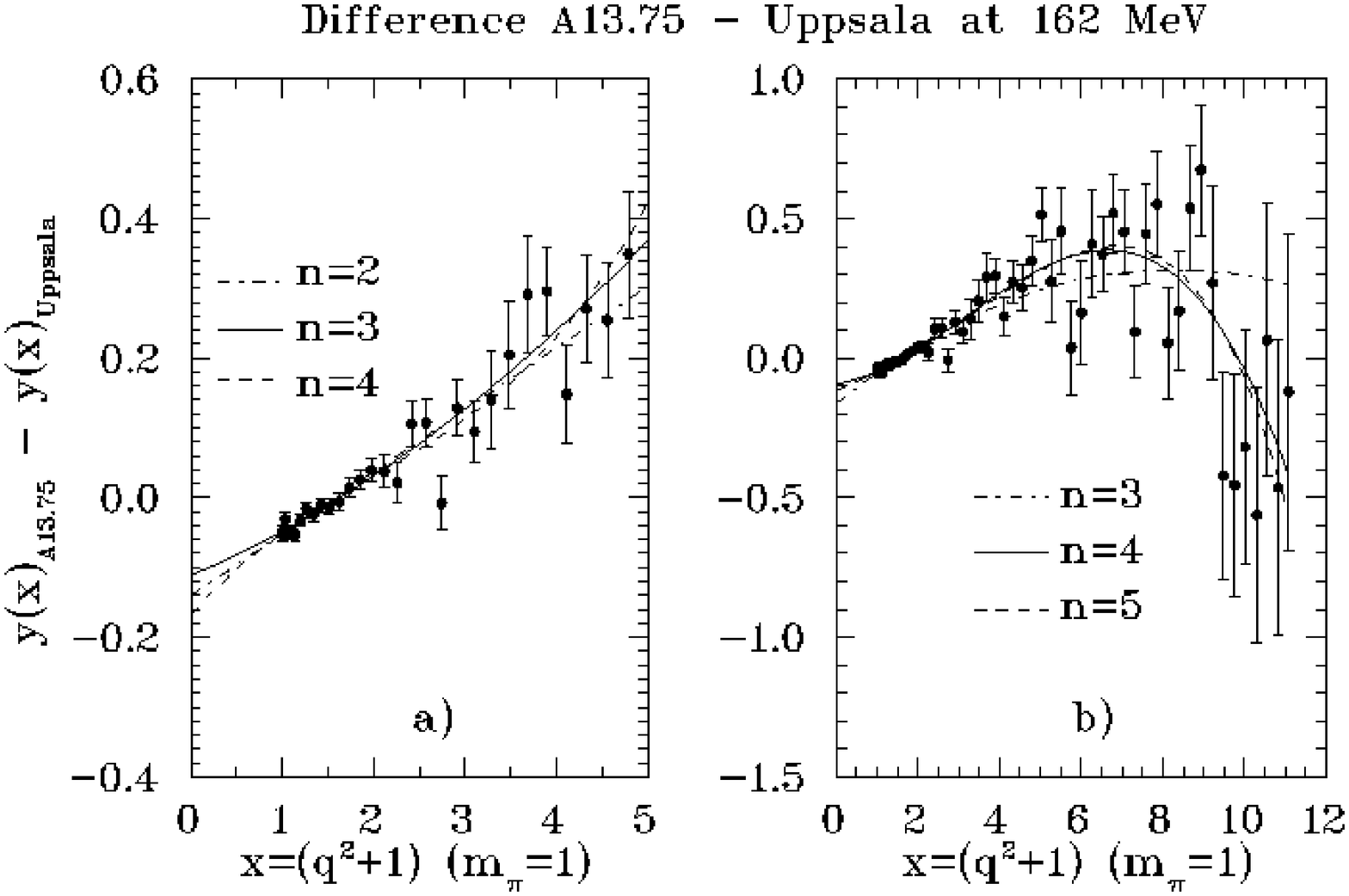}
\caption{As in Fig. \protect\ref{fig:12gda} but for the difference
between model A13.75~\protect\cite{Arndt99b} and Uppsala 
data~\protect\cite{Rahm98,Olsson99}.}
\label{fig:13.75}
\end{center}
\end{figure}

\begin{figure}[htb]
\begin{center}
\includegraphics
[angle=90,width=14cm]{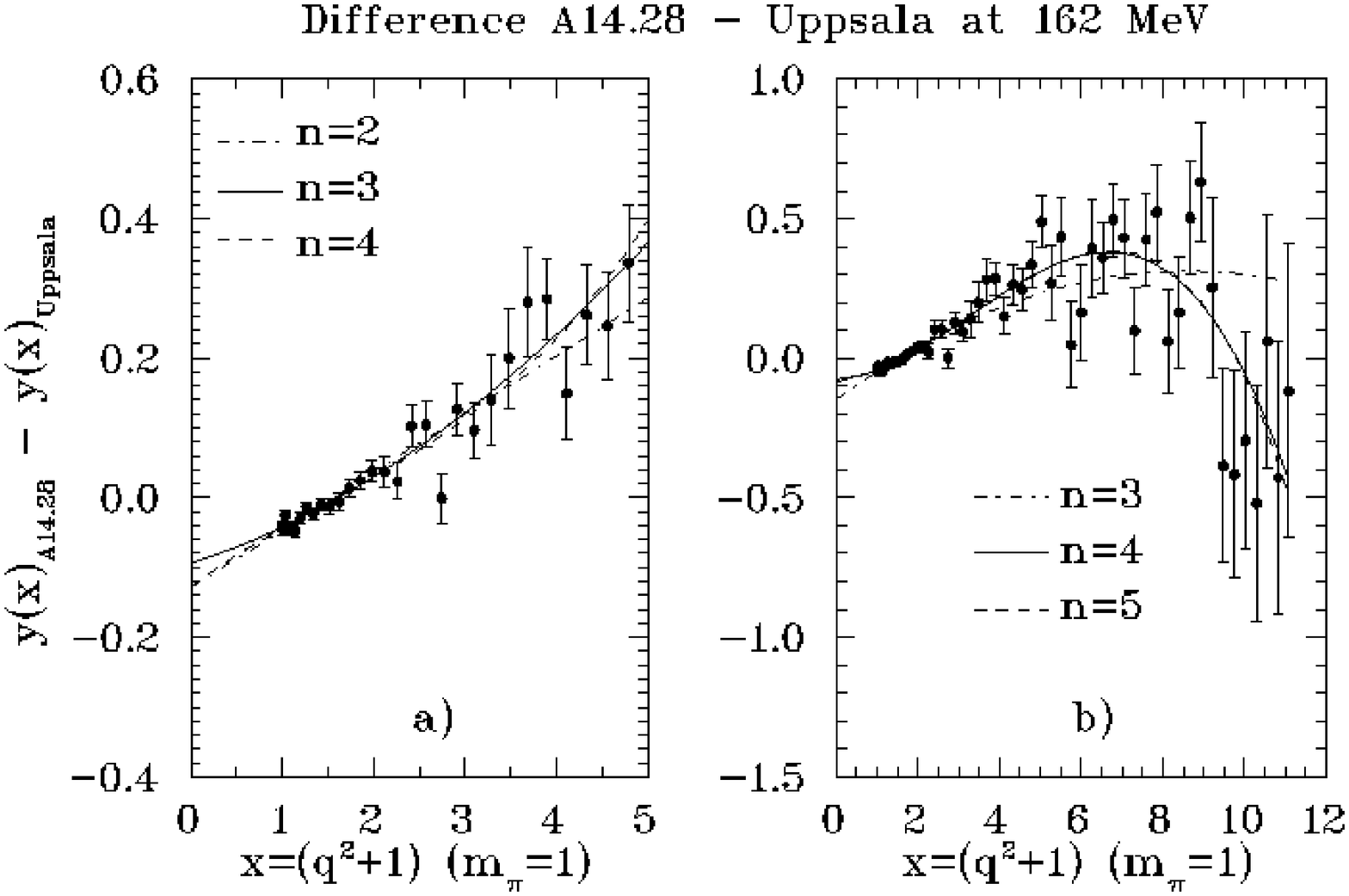}
\caption{As in Fig. \protect\ref{fig:12gda} but for the difference
between model A14.28~\protect\cite{Arndt99b} and Uppsala 
data~\protect\cite{Rahm98,Olsson99}.}
\label{fig:14.28}
\end{center}
\end{figure}

Let us here remind the reader that the difference method analysis is quite 
consistent. One can check, in the Tables~2,~3,~5 
and~6, 
that, at a given $n$ value, the difference between
the results of the difference method 
for the reference models and data is very close to the
systematic shift between the two models. One has,
\begin{eqnarray}
g^{2}_{\pi^\pm}/4\pi(n,\ \mbox{Model A} - \mbox{Uppsala})
& - & g^{2}_{\pi^\pm}/4\pi(n,\ \mbox{Model B} - \mbox{Uppsala})
\nonumber \\
&\simeq&\delta g^{2}_{\pi^\pm}(n,\ \mbox{Model A}-\mbox{Model B}).
	\label{eq:g2syst}
\end{eqnarray}
For instance if we compare in the full range the $n=4$, (A13.75 $-$ Uppsala)
result
to that of (A12.83g $-$ Uppsala) we have a difference
(see Table~6) 
of
0.59~(14.38~-~13.79) which is close to the systematic shift between
these models .65 (see Table~3 
). In Ref.~\cite{Rahm98} where
the comparison models were the Nijmegen potential~\cite{Stoks94}, the
Nijmegen~\cite{Stoks93b} (NI93) and Virginia~\cite{Said,Arndt95}
(SM95) energy dependent PWA's, dispersion of results were smaller.
This could be traced to the fact that these models,
where $g^2_{\pi^\pm}/4\pi$ has been minimised with respect to the $NN$ data,
have a high $q^{2}$ 
momentum more similar to that of the Uppsala data.

\begin{figure}[htb]
\begin{center}
\includegraphics
[angle=90,width=14cm]{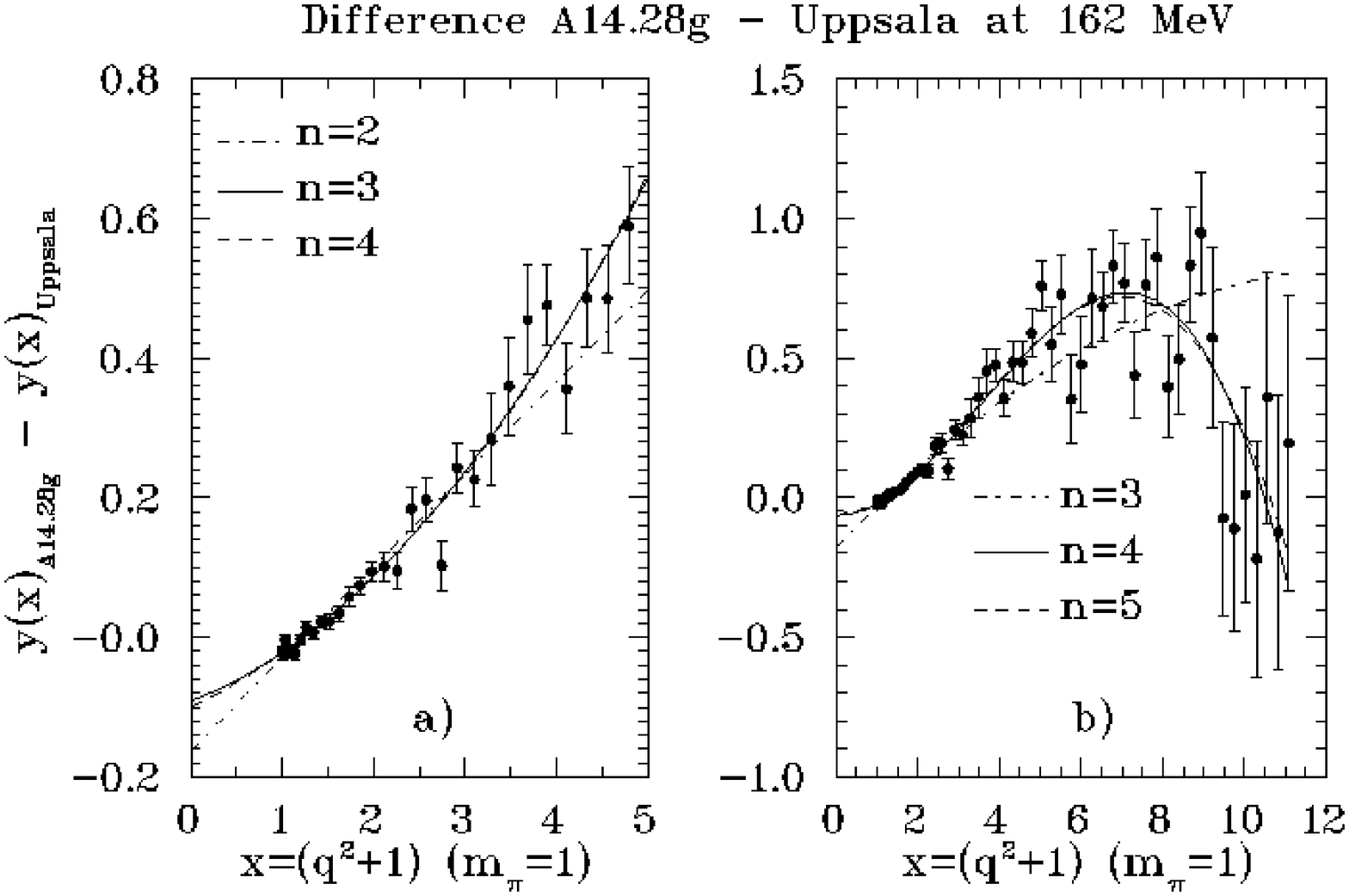}
\caption{As in Fig. \protect\ref{fig:12gda} but for the difference
between model A14.28g~\protect\cite{Arndt99b} and Uppsala 
data~\protect\cite{Rahm98,Olsson99}.}
\label{fig:14.28g}
\end{center}
\end{figure}

Summarising these results we take for $g^{2}$ the following average 
in the reduced range
\begin{eqnarray}
	g^{2}_{\pi^\pm}/4\pi& = &\frac{1}{5}
	[14.69(31)+14.37(31)+14.50(12)+14.96(12)+14.94(12)] \nonumber \\
	& = &14.69(20)
	\label{eq:g2reda}
\end{eqnarray}
It is here only the second value (from A12.83$-$Uppsala) which is slightly 
outside the range of the 2 last values from A14.28 and A14.28g

Taking half of the average of $\vert\delta g^{2}_{\pi^\pm}\vert$ between 
models at $\chi^{2}/N_{df}=1.00$ for the estimation of the 
systematic uncertainty (see Table~5) 
we obtain
\begin{eqnarray}
	g^{2}_{\pi^\pm}/4\pi& = & 14.69\pm 0.20\ \mbox{(stat. + extr.)}
	\pm 0.15\ \mbox{(syst.)}\pm 0.17\ \mbox{(norm.)} \nonumber \\
	& = & 14.69(30)
	\label{eq:g2redb}
\end{eqnarray}
i.e. an {\it accuracy of} 2 \%.
For the full range (see Table~6) 
we have
\begin{eqnarray}
	g^{2}_{\pi^\pm}/4\pi& = &\frac{1}{5}
	[14.41(24)+13.94(25)+14.38(14)+14.83(13)+14.75(13)] \nonumber \\
	& = &14.46(18)
	\label{eq:g2fulla}
\end{eqnarray}
Adding the systematic (estimated as above) and normalisation errors we obtain 
\begin{eqnarray}
		g^{2}_{\pi^\pm}/4\pi& = &14.46\pm 0.18\ \mbox{(stat.+ extr.)}
		\pm 0.15\ \mbox{(syst.)}\pm 0.17\ \mbox{(norm.)}
		\nonumber \\
		& = &14.46(29)
	\label{eq:g2fullb}
\end{eqnarray}
i.e. again an {\it accuracy of} 2\%.
Note that for A12.83 $-$ A14.28 we need to go to $n=7$ to get
$\chi^{2}/N_{df}=1.00$ with a $\delta g^{2}_{\pi^\pm}$ of $-$0.36. The 
value of Eq.~(\ref{eq:g2fullb}}) is to be compared to the value we 
determined in Ref.~\cite{Rahm98}, viz.
\begin{eqnarray}
	g^{2}_{\pi^\pm}/4\pi& = &14.52\pm 0.13\ \mbox{(stat.+ extr.)} 
	\pm 0.15\ \mbox{(syst.)}\pm 0.17\ \mbox{(norm.)} \nonumber \\
	& = &14.52(26) 
	\label{eq:g2prc}
\end{eqnarray}
i.e. an {\it accuracy of} 1.8\%.
It is seen that both determination, Eqs.~(\ref{eq:g2fullb})
and~(\ref{eq:g2prc}), are very close. The determination of
Ref.~\cite{Rahm98}, Eq.~(\ref{eq:g2prc})
 has a 
better statistical extrapolation error which can be understood as the 
comparison model have possibly a better determined high $q^{2}$ 
behaviour as just mentioned above.

\section{Conclusions}
\label{concl}

This analysis of the 162 Mev Uppsala precise experiment on np charge exchange
demonstrates here again that such data can be used for a direct and accurate 
determination of the $\pi N\! N$ coupling constant. We reproduce the
original coupling constant using the present procedures for pseudo-data
from  PWA's as exemplified with the model DA99. The 
extrapolation error  increases with the number of parameters.
 Our value is 7\% larger than the 
Nijmegen~\cite{Stoks93a}
result $g^2_{\pi^\pm} = 13.58 \pm 0.05$, but it is consistent 
with values given in earlier data compilations based on the analysis of 
$\pi N$ and $N\! N$ scattering data~\cite{Dum83}. 

The  data have been used to determine a precise value for the
charged $\pi N\! N$ coupling constant using extrapolation to the pion
pole. Using the most accurate extrapolation method, the Difference
Method, we find $\sqrt{N} g^2_{\pi^\pm} = 14.46 \pm 0.18$
($f^2_{\pi^\pm} = 0.0800 \pm 0.001$) with a systematic error of about
$\pm 0.15$ ($\pm 0.0008$) and a normalisation uncertainty of $\pm 0.17$
($\pm 0.0009$). We do  reproduce the input coupling
constants of models using equivalent pseudo-data. The practical
usefulness of the method, its precision and its relative insensitivity
to systematics appear to be under control.
The pseudo-data demonstrate that considerable precision
is achieved statistically at a single energy. The absolute
normalisation of the data is nevertheless crucial.
The precision of the method  used here has not yet
reached its theoretical limit, but we can point out the key
information necessary for this in the $N\! N$ sector. We need as precise as
possible
unpolarised differential cross sections with an absolute normalisation
of 1 to 2\% to reach a precision of about 1\% in the coupling constant.
For the Uppsala data  an accurate normalisation~\cite{Rahm98,Olsson99} of
2.3 \% was obtained
using integration over the  angular
distribution. It was performed on the part they measured 
(from 72$^\circ$ to 180$^\circ$)
and on the remaining one calculated from PWA's and models. The result was
scaled to the well known experimentally total cross section~\cite{Lis82}.
It is important to extend the angular range of data, 
to be able to achieve an improved normalisation.

For that purpose there is, as we have heard~\cite{Olsson99} in this workshop,
 an experiment in progress at The Svedberg Laboratory. It will
measure the np differential cross section in the forward hemisphere. This
should hopefully determine the normalisation to 1~\% and then lower down the
accuracy on $g^2_{\pi^\pm}/4\pi$ from 1.8~\% to 1.5~\%. 
To the question which was asked
at this workshop ``Does the difference method work at the 1~\% level?''
the answer is so far no, but as we have demonstrate it can work to less than
2~\%, in particular with good reference models we obtained a precision of
1.8~\%. Here with different reference models the precision reached was 2~\%.
Contrary to what was claimed in Ref.~\cite{SWA97} we do not see a large
model dependence, even using ``extreme'' models as the A12.83g
with $g^2_{\pi^\pm}/4\pi$=12.83 or $f^2_c/4\pi$=0.071. The model dependence
is relatively small as can be judged from a comparison of the result here
and that of Ref.~\cite{Rahm98} where while using different reference models
the results agree within less than 0.5~\%.

Let us also mention, as we were told~\cite{Peterson99} in this workshop
that np backward measurement with tagged neutron beams (which leads to
absolute normalisation) are in progress at IUCF for
185 $\ge T_{lab} \le$ 195 MeV
and $90^\circ < \theta < 180^\circ$. If the very backward steeper
shape of
the Uppsala data as compared with earlier
data~\cite{Rahm98,Olsson99,Blomgren99}
is confirmed together with its absolute normalisation then the $\pi NN$
coupling constant cannot be as low as found, for instance, by the Nijmegen
group. If the Uppsala data is correct and if the coupling constant is small
then to reconciliate both, either the total cross section could be off or the
angular distribution in the forward hemisphere could be different of what it
is believed so far from PWA analyses. Hopefully the two above mentioned
experiments in progress should help to clarify the situation.
There will be also 
pp and np
spin-transfer measurements which could  be used to precise the $\pi NN$
coupling constant~\cite{Wissink99}.

In principle, an 
 experiment at one single energy is enough to determine $g^2_{\pi^\pm}/4\pi$, 
for all energies contain similar information. 
Although the method of analysis seems to work well, it is, 
however, useful to deduce the coupling constant from data at several 
energies. This would increase confidence that some unexpected systematic 
effect influences the conclusion.
\vspace{2mm}

We thank the The Svedberg Laboratory crew for the data.
 We are also grateful to J. Blomgren, M. Lacombe and N. Olsson 
 for helpful discussions 
  and to 
W.R. Gibbs for advice on producing pseudo-data from models. TE 
acknowledges an interesting discussion with M. Rentmeester and BL the 
hospitality of the The Svedberg Laboratory.

This work has been financially supported by the Swedish Natural Science
Research Council 
and, gr\^ace au Service Scientifique et Technique de l'Ambassade de France
\`a Stockholm en Su\`ede, par le Minist\`ere des Affaires 
\'Etrang\`eres Fran\c cais.


\pagebreak



\end{document}